\documentclass[]{interact}

\usepackage{epstopdf}

\usepackage[numbers,sort&compress]{natbib}
\bibpunct[, ]{[}{]}{,}{n}{,}{,}
\makeatletter
\def\NAT@def@citea{\def\@citea{\NAT@separator}}
\makeatother

\theoremstyle{plain}

\theoremstyle{definition}

\theoremstyle{remark}

\usepackage{url}
\usepackage{hyperref}
\usepackage{subcaption}
\usepackage{float} 

\begin{document}

\articletype{}

\title{Detecting Faltering Growth in Children via Minimum Random Slopes}

\author{
\name{Jarod Y. L. Lee\textsuperscript{a,b}\thanks{CONTACT Jarod Y. L. Lee. Email: yan.lee@uts.edu.au}, Craig Anderson\textsuperscript{c}, Wai T. Hung\textsuperscript{a,b}, Hon Hwang\textsuperscript{a,b} and Louise M. Ryan\textsuperscript{a,b,d}}
\affil{\textsuperscript{a}School of Mathematical and Physical Sciences, University of Technology Sydney, Sydney, Australia;
\textsuperscript{b}Australian Research Council Centre of Excellence for Mathematical \& Statistical Frontiers;
\textsuperscript{c}School of Mathematics \& Statistics, University of Glasgow, UK;
\textsuperscript{d}Harvard T. H. Chan School of Public Health, Harvard University, Cambridge, USA}
}

\maketitle

\begin{abstract}
A child is considered to have faltered growth when increases in their height or weight starts to decline relative to a suitable comparison population. However, there is currently a lack of consensus on both the choice of anthropometric indexes for characterizing growth over time and the operational definition of faltering. Cole's classic conditional standard deviation scores is a popular metric but can be problematic, since it only utilizes two data points and relies on having complete data. In the existing literature, arbitrary thresholds are often used to define faltering, which may not be appropriate for all populations. In this article, we propose to assess faltering via minimum random slopes (MRS) derived from a piecewise linear mixed model. When used in conjunction with mixture model-based classification, MRS provides a viable method for identifying children that have faltered, without being dependent upon arbitrary standards. We illustrate our work via a simulation study and apply it to a case study based on a birth cohort within the Healthy Birth, Growth and Development knowledge integration (HBGDki) project funded by the Bill and Melinda Gates Foundation. 
\end{abstract}


\section{Introduction}

Faltering growth is associated with a range of health conditions such as postnatal depression and diarrhea \cite{Assis2005Growth, Hung2018on, OBrien2004Postnatal}. As such, there is substantial interest from epidemiologists in identifying children that experienced faltering growth, since this is imperative in designing prospective research studies. For instance, investigation on children who have faltered may reveal that many of them took a particular type of dietary supplement. Studies can then be designed to investigate this hypothesis, and if significant evidence is present then action can be taken across the cohort. When it comes to individual child, it is also important for paediatrician to identify those who have faltered in order to facilitate timely intervention for improving long-term health outcomes.


While there is no agreed definition among practitioners \cite{Wilcox1989Failure, Olsen2006Failure}, faltering is commonly regarded as slower than expected rate of growth when compared to a suitable population. To accurately identify children who have faltered, it is important to formulate models that accurately capture the rate of a child's growth, also known as growth velocity. Classification rules can then be applied to the estimated velocities to separate children who have faltered from those who have not.

There exist a variety of statistical models to estimate growth velocities, the most common being the standard deviation score (SDS), which is the change in relative height or weight between two given time points. Although SDS is convenient and easy to implement, it comes with several limitations since it only utilizes two data points and relies on having complete data.  Note that the heights and weights are standardized according to the WHO's z-scores \cite{WHO2006}, which are computed to reflect the global ``healthy" average, allowing for comparison across populations of a given age and gender. The standardization gives rise to the height-for-age z-scores (HAZ) and weight-for-age z-scores (WAZ), respectively. Thus, a value of $1$ on the HAZ scale corresponds to a child who is one standard deviation above the mean of the WHO reference population for the same age and gender. If the comparison population is appropriate, we expect the z-scores to have a mean of zero at any combination of age and gender.

The random slopes (RS) metric circumvents some of the limitations of SDS. It is based on a linear mixed model, thus allowing the use of more than two data points while taking into account of the child-specific departures from the global trajectory. However, the linearity assumption of the approach is overly restrictive in capturing the true growth trajectory, and there is a need for more flexible models. Anderson et al. \cite{Anderson2018comparing} compared various growth models fitted on both raw and standardized measurements. They concluded that fitting a piecewise linear mixed model on standardized data provides a robust approach for accurate estimation and prediction for a wide range of child growth data.  Consequently, the piecewise linear mixed model is used as the basis for deriving two new velocity metrics in this article. 


Once we have estimated the growth velocity for each child based on their fitted trajectories, the next step is to classify the children into faltering and non-faltering groups based on the magnitude of their velocity. Intuitively, children with the lowest growth velocity estimates should be classified as having faltered. In the existing literature, threshold-based classification is often used to define the faltering group, where the threshold is chosen arbitrarily \cite{Gonzalez2017faltering, hall2000growth, Leung2017conditional, OBrien2004Postnatal, ud2013Growth}. For example, ud Din et al. \cite{ud2013Growth} and Leung et al. \cite{Leung2017conditional} defined abnormal growth as velocity estimates below the metric-specific $5$th and $10$th percentile, respectively.  Threshold-based classification is ideal if we know the true proportion of faltering children, but this is not the case in practice. In particular, the threshold is not constant across different birth cohorts, even for cohorts within the same geographical region across different years.


The purpose of this article is twofold. Firstly, we propose two new velocity metrics; the average random slopes (ARS) and the minimum random slopes (MRS), both based on a piecewise linear mixed model. Some authors have proposed the use of conditional models, which adjust for a regression to the mean effect by considering the distribution of baseline measurements. \cite{Cole1994growth, Cole1995conditional, Cole1998presenting, Leung2017conditional}. In this article, we consider both unconditional and conditional models. Secondly, we propose a mixture model-based classification to classify children into faltering and non-faltering groups. Our simulation studies show that the classification performance of MRS, when used in conjunction with mixture model-based classification, is superior to the other velocity metrics under consideration. In particular, the classification results based on mixture models are comparable to that of threshold-based classification when using the true threshold.

The remainder of the article is organized as follows. Section \ref{sec:velmetric} describes the various velocity metrics considered in this article. In Section \ref{sec:classification}, we outline the mixture model-based classification. An account of the statistical packages used to implement the methods discussed in this article is provided in Section \ref{sec:software}. In Section \ref{sec:simulation}, we compare the performance of our proposed velocity metrics and mixture model-based classification with existing methods via simulation studies. We then apply the methods to the India data set obtained from the Healthy Birth, Growth and Development knowledge integration (HBGDki) project, a collaboration funded by the Bill and Melinda Gates Foundation. In the concluding section, we discuss the limitations of the proposed methods and provide suggestions for future research.

\section{Velocity metrics} \label{sec:velmetric}
In this section, we describe the various velocity metrics, including the proposed average and minimum random slopes. 

\subsection{Standard deviation scores (SDS)}

SDS is perhaps the most commonly used velocity metric in child growth literature, because it is straightforward to calculate and easy to interpret. The classical SDS is given by:
\begin{equation}
SDS_{i} = z_{i1} - z_{i0},
\end{equation}
where $z_{i1}$ and  $z_{i0}$ are the z-scores for child $i$ at the end and the beginning of a pre-determined time interval, respectively.  The growth velocity for child $i$ is given by $SDS_{i}$.

Cole \cite{Cole1994growth, Cole1995conditional, Cole1998presenting} also suggested a variation of the SDS metric to account for regression to the mean (RTM) .  The classical formulation of conditional SDS (cSDS) is given by:
\begin{equation}
cSDS_i = \frac{z_{i1}-r z_{i0}}{\sqrt{1-r^2}},
\end{equation}
where $r$ is the Pearson's correlation between the z-scores at time $0$ and time $1$. The growth velocity for child $i$ is given by $cSDS_{i}$, which can also be derived from a linear regression model of $z_{i1}$ on $z_{i0}$  \cite{Cole1994growth}. This forms the basis for considering subsequent velocity metrics that are derived from regression-based models.


SDS is simple to implement, but it does rely on the children having complete data. Thus children with only one observation have to be discarded from analysis. Moreover, the classical formulation of SDS can only accommodate two measurements, resulting in a loss of information as additional data cannot be utilized.  Since SDS is based solely on child-specific measurements, it is more susceptible to measurement errors compared to velocity metrics derived from random effects models which ``borrow strength" from the other children.  In particular, SDS is calculated from the difference between two measurements, thereby combining the imprecision of the two readings which may or may not cancel out each other.



\subsection{Random slopes (RS)}
The conditional SDS metric is derived from a regression setting, so it is natural to consider extensions using other regression approaches. One such extensions is the RS metric, based on a linear mixed model, which can account for the dependency structure inherent within multiple measurements via the incorporation of random effects (left-hand side of Figure \ref{fig:kid1}) \cite{Laird1982random}. The underlying model is given by:
\begin{equation}
z_{ij} = \omega_{i0} + \omega_{i1} t_{ij} + \epsilon_{ij},
\end{equation}
where $z_{ij}$ denotes the z-score for the $i^{th}$ subject on the $j^{th}$ measurement occasion at time $t_{ij}$. Here, $\omega_{i0}$ and $\omega_{i1}$ are the random effects for child $i$, which are assumed to come from a multivariate Gaussian distribution with mean vector $(\beta_0, \beta_1)$ and unstructured covariance matrix $\Sigma$.  The random effects capture each child's deviation from the global trajectory that is characterized by the mean vector. The error term $\epsilon_{ij}$ is assumed to be independent and  drawn from a Gaussian distribution with mean $0$ and variance $\sigma^2$. The growth velocity for child $i$ is given by  $\omega_{i1}$.


To adjust for RTM, Leung  et al. \cite{Leung2017conditional} proposed the conditional RS (cRS) metric, which extends the linear mixed model by including the baseline-age interaction term as a predictor:
\begin{equation}
z_{ij} = \omega_{i0} + \omega_{i1} t_{ij} + \beta_2 t_{ij} z_{i0} + \epsilon_{ij}.
\end{equation}
Here, $z_{i0}$ is the baseline z-score derived from the first observation of the interval under consideration, which is treated as a predictor instead of a response. Following Leung et al. \cite{Leung2017conditional}, we excluded the main effect of $z_{i0}$ from the model as the variability in baseline measurements is already captured by $\omega_{i0}$.  Using this approach, the growth velocity can be calculated by dividing the difference in the predicted values between two data points by the associated time interval. In the special case where the covariance matrix is diagonal, ie. there is no correlation between $\omega_{i0}$ and $\omega_{i1}$, the conditional RS metric derived from this formulation can be viewed as a generalization of the classical cSDS metric \cite{Leung2017conditional}. 

RS allows the use of more than two data points for estimation and does not require the children to be measured at the same time point. The ability of the random effects to borrow strength across children is an added advantage since it allows for velocity estimation for children with at least one observation.  Since RS is derived based on a random effects model, it is less prone to measurement errors.  That being said, the inherent linearity structure imposed by the model is overly restrictive and may not allow the true shape of the growth trajectories to be captured.

\subsection{Average random slopes (ARS)}
ARS is based on the piecewise linear mixed model, also known as the broken stick or linear spline model with random effects (right-hand side of Figure \ref{fig:kid1}) \cite{Howe2016linear, Meyer2005random}.  The piecewise linear mixed model extends the linear mixed model, where a series of straight lines are fitted to segments of the predictor separated by a set of predefined  region boundaries, also known as knots, with the constraint that the fitted curve must be continuous. This results in a flexible class of models that are capable of capturing the non-linear structure in the data. Changing the number and location of the knots allows flexibility over regions where the curve changes rapidly and prevents overfitting over regions where the curve appears to be more stable. There exist several methods for selecting the number and location of the knots \cite[p. 1860]{Howe2016linear}. In a simulation study, Anderson et al. \cite{Anderson2018comparing} investigated the effect of the number of knots on the mean squared error (MSE) of prediction. They found that MSE remained fairly stable regardless of the number of knots for models with 4 or more knots. In this article, we consider evenly spaced knots as the curve is fitted on the standardized measurements. Equally spaced knots are also important for the ARS method so that we are not averaging segments of completely different lengths. Likewise, for the MRS method to be discussed in the next section, it would be unfair to compare a segments covering different lengths, for example comparing a segment covering 0.5 years to that of 0.01 years.

Mathematically, a piecewise linear mixed model with $K$ internal knots placed at $K+1$ ordered ages $\kappa_0 = \kappa_1 < \dots < \kappa_K < \kappa_{K+1}$ can be represented as
\begin{equation}
z_{ij} = \sum_{k=1}^{K+1} \omega_{ik} t^B_{ik} + \epsilon_{ij},
\end{equation}
where $t^B_{ik}$ is the B-spline (``B" stands for basis \cite{de1978practical}) transformation of $t_{ij}$ and $\omega_{ik}$ is the child-specific random effects for the $k^{th}$ B-spline, with mean $\beta_{k}$ and variance that is correlated across $k$. The B-spline transformation yields an equivalent fit to the more interpretable truncated basis function, but is known to have more stable numerical properties \cite[p. 70]{ruppert2003semiparametric}. The left boundary knot $\kappa_0$ is conveniently set to the minimum age and is equal to $\kappa_1$, whereas the right boundary knot $\kappa_{K+1}$ is set to the maximum age \cite[Sec. 3: model formulation]{brokenstick}. This formulation implies that the minimum age is one of the internal knots by definition. The growth velocity for child $i$ is given by the slopes weighted by their respective interval duration across the segments. More conveniently, the velocity can also be obtained by dividing the difference between the child's predicted values at $\kappa_{K+1}$ and $\kappa_0$ by their interval duration. 

To account for RTM, we can extend the piecewise linear mixed model by including a baseline-age interaction term:
\begin{equation}
z_{ij} = \sum_{k=1}^{K+1} \omega_{ik} t^B_{ik} + \beta_{K+2} t_{ij} z_{i0} + \epsilon_{ij}.
\end{equation}
The resulting conditional ARS (cARS) metric can be derived similarly as in ARS, but using predicted values from the updated model.

\subsection{Minimum random slopes (MRS)}
MRS is based on the same model as ARS, that is,  the piecewise linear mixed model. However, instead of weighting the slopes across the different segments of the predictor by their associated interval, we choose the minimum of the slopes across the segments.  The idea is that if we want to identify children who have faltered within a given period of time, we expect a classifier built based on the minimum velocities to perform well since the metric is not ``contaminated" by values from the other segments. The conditional MRS (cMRS) that take into account of RTM can be derived in a similar way as MRS, but using predicted values from the piecewise linear mixed model that includes the baseline-age interaction term as a predictor.

\subsection{Example: velocity calculation}

\begin{figure}[htb]
	\centering
	\includegraphics[width=\textwidth]{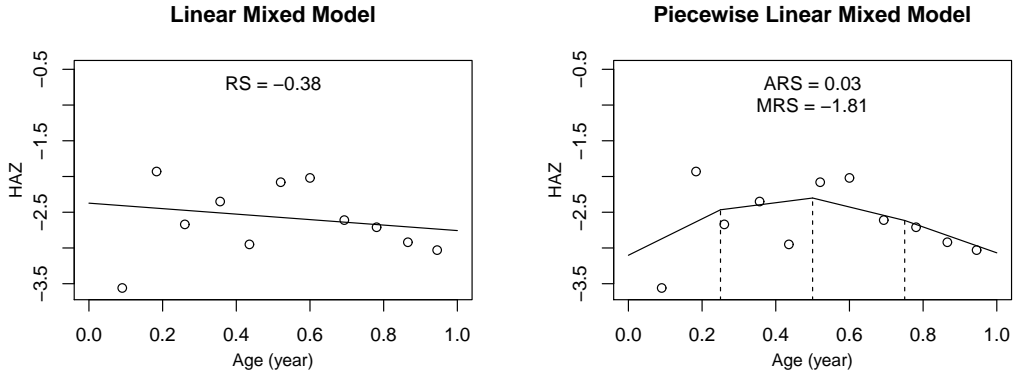} 
	\caption{Velocity estimates for a single child within the India data set. From left to right, random slopes (RS) based on the linear mixed model; average random slopes (ARS) and minimum random slopes (MRS) based on the piecewise linear mixed model with internal knots placed at 0, 0.25, 0.5, and 0.75 years.} \label{fig:kid1}
\end{figure}

Figure \ref{fig:kid1} presents the fitted trajectories of a randomly chosen child from the India data set, along the various velocity estimates. The data set will be discussed in detail in Section \ref{sec:application}. We are interested in estimating the growth velocity of this particular child across the first year, with the aim of determining whether the child has faltered during this period of time.  For the purpose of illustration, we only consider unconditional velocity metrics in this section. 

The nearest HAZ to age $0$ was observed at age $0.09$ with a value of $-3.56$, and the nearest HAZ to age $1$ was observed at age $0.95$ with a value of $-3.03$. Thus, the SDS for this child across the first year is $-3.03 -(-3.56) = 0.53$. Under the linear mixed model, the fixed slope estimate is $-1.16$ and the predicted random slope for this child is $0.78$, giving rise to the RS estimate of $-1.16 + 0.78 = -0.38$, which is the child-specific slope. The piecewise linear mixed model provides a more realistic fit, with predicted values at age $0$, $0.25$, $0.5$, $0.75$, and $1$ given by $-3.102$, $-2.466$, $-2.301$, $-2.615$, and $-3.068$, respectively. The velocity across the first segment of age bounded by $0$ and $0.25$ is given by $\lbrace-2.466 - (-3.102)\rbrace / 0.25 = 2.544$. Likewise, the velocities across the segments of age $0.25$ to $0.5$, $0.5$ to $0.75$, and $0.75$ to $1$  are $0.66$, $-1.256$, and $-1.812$, respectively. The ARS estimate across the first year is thus $0.25*(2.544 + 0.66 - 1.256 - 1.812) = 0.034$. More conveniently, the ARS can be calculated by first taking the difference between the predicted values at the end and the beginning of the interval, and then dividing by the associated interval duration, that is, $\lbrace -3.068 - (-3.102) \rbrace / 1 = 0.034$. The MRS estimate is given by the minimum of the velocities across the segments of age, which is $-1.812$.  Visually, the MRS metric is able to reflect the period of decline towards the end of the first year.



\section{Mixture model-based (MM) classification} \label{sec:classification}

We propose using a Gaussian mixture model to identify potential heterogeneous components in the estimated velocities. Our objective is to classify the children into faltering and non-faltering groups based on the estimated velocities, hence we fix the number of components to be two. Let $x_i$ be the estimated velocity for child $i$. Under a two-component Gaussian mixture model, the density of $x_i$ is
\begin{equation}
	f(x_i) = \sum_{k=1}^{2} \lambda_k \phi_k(x_i),
\end{equation}
where $\phi_k$ are drawn from Gaussian density function with mean $\mu_k$ and standard deviation $\sigma_k$, and $\lambda_k$ are positive weights which sum to unity \cite{mixtools}. By definition, the component with the lower mean is regarded as the faltering group, and the component with the larger mean is the non-faltering group. Children are assigned into one of the groups based on their ``posterior" probability, that is, the probability that the observation comes from each of the two components given the velocity and parameter estimates. We use 0.5 as the cutoff value. Intuitively, we expect MRS-based velocity estimates to be more heterogeneous compared to RS and ARS-based estimates, rendering the Gaussian mixture model a suitable method for separating the velocity estimates into two different components.



\section{A note on the software packages used} \label{sec:software}
The simulations and analyses were carried out using the R statistical computing environment, version 3.4.3 \cite{R}. The linear and piecewise linear mixed models were fitted using the lmer() function within the ``lme4" package \cite{lme4}.  The piecewise linear mixed model were fitted by first generating a B-spline basis matrix of degree 1 for the age variable using the bs() function within the ``splines" package, and then feeding the matrix into the lmer() function as predictors. The piecewise linear mixed model can also be fitted using the brokenstick() function within the ``brokenstick" package \cite{brokenstick}, which at the time of writing is only able to accommodate a single covariate. Fitting the piecewise linear mixed model using the lmer() function allows us to incorporate additional predictors such as the baseline-age interaction, which is needed to adjust for the regression to the mean effect. The Cohen's Kappa statistics and its associated p-value in the simulation were calculated using the kappa2() function within the ``irr" package \cite{irr}. The Gaussian mixture model were fitted using the normalmixEM() function within the ``mixtools" package \cite{mixtools}.


\section{Simulation study} \label{sec:simulation}

\subsection{Simulation setup}
We generated a population of $1000$ children and assigned each child with a unique growth trajectory.  The trajectories for the general population are generated from a linear mixed model of the form 
\begin{equation}
	z_{ij} = \omega_{i} t_{ij} + \epsilon_{ij},
\end{equation}
where $z_{ij}$ denotes the z-scores for the $i^{th}$ child on the $j^{th}$ measurement occasion at time $t_{ij}$, in years. Here, $\omega_{i}$ are the random slopes, which were independently drawn from Gaussian(-1, $\sigma_\omega$), independent of the errors $\epsilon_{ij}$, which were independently drawn from Gaussian(0, $\sigma_{\epsilon}$). The first and second arguments of the Gaussian distribution represent the mean and standard deviation, respectively. We set $\sigma_{\omega}$ and $\sigma_{\epsilon}$ to be 0.25 and 0.3, respectively. This setup is in line with the growth pattern of children in low-to-middle income countries, in which the z-scores decline throughout the first year. The intercept term was deliberately omitted to remove the effect of regression to the mean. 

\begin{figure}[htb]
	\centering
	\includegraphics[width=0.8\textwidth]{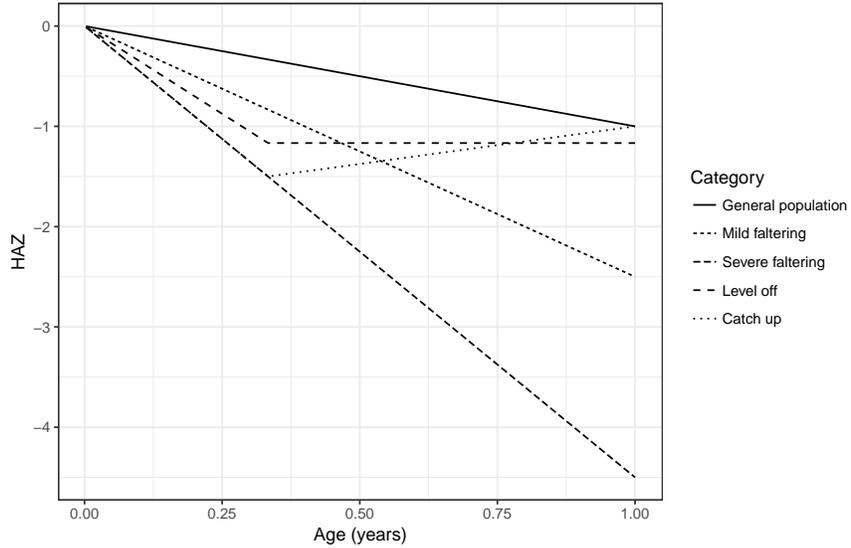} 
	\caption{Averaged trajectories of the simulated population subgroups. Each child's trajectory deviates from the population average via a child-specific random slope. The intercept was set to $0$ to remove the effect of regression to the mean.} \label{fig:aveTraj}
\end{figure}

We selected a subset of children on which to impose a faltering episode.  Four subgroups were created within this subset to represent a variety of faltering scenarios (Figure \ref{fig:aveTraj}). In practice, there will be more faltering scenarios that we could possibly consider in the simulation. The patterns that we chose represent the ones that are common across the datasets within the Healthy Birth, Growth and Development knowledge integration (HBGDki) project database.

The first two subgroups were created by setting the $\omega_{i}$ values as follow:
\begin{itemize}
	\item[(i)] Mild faltering :  $\omega_{i}$ were drawn from Gaussian(-2.5, $\sigma_{\omega}$). This results in an average velocity of $-2.5$ across this subgroup. 
	\item[(ii)] Severe faltering:  $\omega_{i}$ were drawn from Gaussian(-4.5, $\sigma_{\omega}$). This results in an average velocity of $-4.5$ across this subgroup. 
\end{itemize}

The remaining two subgroups were created by imposing a change in slopes at $\kappa = 1/3$ years while maintaining continuity. This is done via:
\begin{equation}
z_{ij}=\begin{cases}
\omega_{i1} t_{ij} + \epsilon_{ij}, & \text{if $t_{ij}<\kappa$}.\\
\omega_{i1} \kappa + \omega_{i2} (t_{ij}-\kappa) + \epsilon_{ij}, & \text{otherwise}.
\end{cases}
\end{equation}
The error term $\epsilon_{ij}$ is defined as above and $\omega_{i2}$ can be interpreted as the change in slope between $t_{ij} < \kappa$ and $t_{ij} > \kappa$ for child $i$, and $\omega_{i1} + \omega_{i2}$ is the change in $z_{ij}$ for child $i$ for a unit increase in $t_{ij}$, when $t_{ij} > \kappa$. The $\omega_{i1}$ and $\omega_{i2}$ values were set according to the following:
\begin{itemize}
	\item[(iii)] Level off: $\omega_{i1}$ and $\omega_{i2}$ were drawn from Gaussian(-3.5, $\sigma_{\omega}$) and Gaussian(0, $\sigma_{\omega}$), respectively. This results in an average velocity of $-7/6$ across this subgroup. 
	\item[(iv)] Catch up: $\omega_{i1}$ and $\omega_{i2}$ were drawn from Gaussian(-4.5, $\sigma_{\omega}$) and Gaussian(0.75, $\sigma_{\omega}$), respectively. This results in an average velocity of $-1$ across this subgroup. 
\end{itemize}

For each children, we sampled 6 to 12 observations from their simulated trajectories. To be ``fair" to the SDS metric, one observation was sampled at an age between 0 and 1/12 years old and another between 11/12 and 1 year old. The remaining observations were sampled in the period between 1/12 and 11/12 years old. 

We generated populations where the percentages of faltered children were 5\%, 10\%, and 20\%. For each simulation scenario, 100 iterations were run, and the results were averaged. We compared the threshold-based (TH) and mixture model-based (MM) classifications under the following velocity metrics: SDS, RS, ARS, and MRS. For threshold-based classification, the thresholds were selected based on the true proportions of faltering children. For instance, we used a threshold of 10\% for a population with 10\% of faltering children.  The piecewise linear mixed model in which ARS and MRS are based on was fitted with internal knots at $0$, $0.25$, $0.5$, and $0.75$ years old.

\subsection{Results}

Table \ref{tab:sim_true+} shows the number of true positives for TH and MM classification built using the various velocity metrics. In our context, true positives translates to the number of children who actually belong in the faltering subgroups among those who were classified as faltered. In the case of severe faltering, all the methods performed similarly well in picking up the faltering children, even for simple method such as SDS. However, SDS did not perform well in situations involving more complicated patterns such as ``level off" and ``catchup". The classification performances based on RS and ARS were comparable, suggesting that these metrics are exchangeable in practice. Classification based on MRS was most accurate in terms of the ability to pick up more complex patterns. Overall, there was little difference between TH and MM classification under the MRS metric. 

Table \ref{tab:sim_agree} shows the agreement between TH and MM classification under the different velocity metrics. A strong agreement is indicated by a low percentage discordance and a high value of Cohen's kappa. Although the significance tests for Cohen's kappa were all significant at the $1\%$ level, the two classification methods based on MRS agreed the most. Given that the results of TH classification were obtained using the true threshold, MM classification based on MRS proved to be a competitive alternative in real life situation where the true threshold is unknown.


\subsection{Changing experimental design: sparse data}

Tables \ref{tab:sim_true+sparse} and \ref{tab:sim_agreesparse} shows the true positives and agreement between TH and MM classification under the data-deficient scenario, when the number of observations per children were decreased from between 6 and 12 observations, to between 2 and 6 observations. As expected, all the methods were able to pick up the cases of severe faltering. When the true proportion of faltering children is small, the performance of MRS was similar to that of RS and ARS in terms of picking up children who have faltered. However, as the true proportion of faltering children increases, MRS outperformed both RS and ARS. It remained true that both TH and MM classification methods using MRS agreed the most compared to other velocity metrics.

\begin{table}[hp]
	\caption{True positives of threshold-based (TH) and mixture model-based (MM) classification under various velocity metrics. The numbers are averages across 100 simulations.}
	\label{tab:sim_true+}
	\subcaption{Proportion faltering = 5\%.}
	\begin{tabular}{lccccccccc} 
		\toprule
		&  & \multicolumn{2}{c}{SDS} & \multicolumn{2}{c}{RS}  & \multicolumn{2}{c}{ARS} & \multicolumn{2}{c}{MRS} \\    \cmidrule(){3-4} \cmidrule(l){5-6} \cmidrule(l){7-8} \cmidrule(l){9-10}
		& N   & TH  & MM & TH & MM  & TH & MM & TH & MM \\ \midrule
		Mild               & 25  & 19.67 & 11.20 & 24.83 & 24.28   & 24.83   & 24.26   & 24.43 & 24.36 \\
		Severe           & 10   & 10.00 & 10.00 & 10.00 & 10.00   & 10.00   & 10.00   & 10.00  &  10.00 \\
		Level              & 10  & 0.30   & 0.05  & 3.33   & 1.28     & 2.99    & 1.03     & 5.98   & 5.64 \\ 
		Catchup         & 5   & 0.02   & 0.00  & 1.94    & 0.86    & 1.71     & 0.64     & 3.93   &  3.75 \\ \midrule
		\textbf{Total} & 50  & 29.99 & 21.25 & 40.10 & 36.42  & 39.53  & 35.93    & 44.34 &  43.75 \\ \bottomrule
	\end{tabular}
	\vskip5mm
	
	\subcaption{Proportion faltering = 10\%.}
	\begin{tabular}{lccccccccc} 
		\toprule
		&  & \multicolumn{2}{c}{SDS} & \multicolumn{2}{c}{RS}  & \multicolumn{2}{c}{ARS} & \multicolumn{2}{c}{MRS} \\    \cmidrule(){3-4} \cmidrule(l){5-6} \cmidrule(l){7-8} \cmidrule(l){9-10}
		& N   & TH  & MM & TH & MM  & TH & MM & TH & MM \\ \midrule
		Mild               & 50  & 42.54  & 26.99 & 49.81 & 48.98  & 49.81   & 48.89   & 48.41 & 48.93 \\
		Severe           & 20  & 20.00  & 19.99 & 20.00 & 20.00  & 20.00   & 20.00   & 20.00 & 20.00 \\
		Level              & 20  & 0.94    & 0.08  & 7.23   & 2.36    & 6.75     & 1.94     & 15.38 & 15.87 \\ 
		Catchup         & 10   & 0.21    & 0.05  & 4.13   & 1.55    & 3.67     & 1.12     & 9.25 & 9.26  \\ \midrule
		\textbf{Total} & 100  & 63.69 & 47.11 & 81.17  & 72.89  & 80.23  & 71.95    & 93.04 & 94.06 \\ \bottomrule
	\end{tabular}
	\vskip5mm
	
	\subcaption{Proportion faltering = 20\%.}
	\begin{tabular}{lccccccccc} 
		\toprule
		&  & \multicolumn{2}{c}{SDS} & \multicolumn{2}{c}{RS}  & \multicolumn{2}{c}{ARS} & \multicolumn{2}{c}{MRS} \\    \cmidrule(){3-4} \cmidrule(l){5-6} \cmidrule(l){7-8} \cmidrule(l){9-10}
		& N   & TH  & MM & TH & MM  & TH & MM & TH & MM \\ \midrule
		Mild               & 100  & 92.18  & 68.81   & 99.85  & 98.77    & 99.88   & 98.70   & 96.43 & 98.02 \\
		Severe           & 40   & 40.00  & 40.00   & 40.00  & 40.00   & 40.00   & 40.00   & 40.00 & 39.60 \\
		Level              & 40   & 3.95    & 0.51     & 15.47   & 4.84     & 14.46   & 3.79      & 35.13 & 36.98 \\ 
		Catchup         & 20   & 1.08    & 0.13     & 7.60     & 2.47     & 6.70     & 1.56      & 19.40 & 19.47 \\ \midrule
		\textbf{Total} & 200  & 137.21 & 109.45 & 162.92 & 146.08 & 161.04 & 144.05   & 190.96 & 194.07 \\ \bottomrule
	\end{tabular}
\end{table}

\begin{table}[h]
	\tbl{Agreement between threshold-based (TH) and mixture model-based (MM) classification under various velocity metrics, measured using percentage discordance (\%D), Cohen's kappa ($\kappa$) and its significance test (significance level = $1\%$). The numbers are averages across 100 simulations.}
	{\begin{tabular}{lccccccccc} \toprule
			 &   \multicolumn{3}{c}{5\%}  &  \multicolumn{3}{c}{10\%} &  \multicolumn{3}{c}{20\%} \\     \cmidrule(){2-4}  \cmidrule(l){5-7} \cmidrule(l){8-10} 
			Metric  & \%D & $\kappa$ & \%Sig  & \%D & $\kappa$ & \%Sig  & \%D & $\kappa$ & \%Sig \\ \midrule
			SDS & 2.72 & 0.61 & 100 & 4.89 & 0.65 & 100 & 8.25 & 0.69 & 100 \\
			RS   & 1.21  & 0.86 & 100 & 2.50 & 0.84 & 100 & 4.94 & 0.83 & 100 \\
			ARS & 1.26  & 0.85 & 100 & 2.58 & 0.84 & 100 & 5.16 & 0.82 & 100 \\ 
			MRS & 0.52 & 0.94 & 100 & 0.65 & 0.96 & 100 & 2.31 & 0.93 & 100 \\ \bottomrule
	\end{tabular}}
	\label{tab:sim_agree}
\end{table}

\begin{table}[hp]
	\centering
	\caption{True positives of threshold-based (TH) and mixture model-based (MM) classification under various velocity metrics. The numbers are averages across 100 simulations.}
	\label{tab:sim_true+sparse}
	\subcaption{Proportion faltering = 5\%.}
	\begin{tabular}{lccccccccc} 
		\toprule
		&  & \multicolumn{2}{c}{SDS} & \multicolumn{2}{c}{RS}  & \multicolumn{2}{c}{ARS} & \multicolumn{2}{c}{MRS} \\    \cmidrule(){3-4} \cmidrule(l){5-6} \cmidrule(l){7-8} \cmidrule(l){9-10}
		& N   & TH  & MM & TH & MM  & TH & MM & TH & MM \\ \midrule
		Mild               & 25  & 19.86 & 12.00 & 24.40 & 22.16   & 24.39   & 22.15   & 24.03 & 22.07 \\
		Severe           & 10   & 10.00 & 9.99 & 10.00 & 10.00   & 10.00   & 10.00   & 10.00  &  10.00 \\
		Level              & 10  & 0.39   & 0.06  & 1.51   & 0.41     & 1.39    & 0.37    & 2.62   & 1.64 \\ 
		Catchup         & 5   & 0.11   & 0.01 & 0.84    & 0.22    & 0.72     & 0.21     & 1.83   &  1.15 \\ \midrule
		\textbf{Total} & 50  & 30.36 & 22.06 & 36.75 & 32.79  & 36.50  & 32.73    & 38.48 &  34.86 \\ \bottomrule
	\end{tabular}
	\vskip5mm
	
	\subcaption{Proportion faltering = 10\%.}
	\begin{tabular}{lccccccccc} 
		\toprule
		&  & \multicolumn{2}{c}{SDS} & \multicolumn{2}{c}{RS}  & \multicolumn{2}{c}{ARS} & \multicolumn{2}{c}{MRS} \\    \cmidrule(){3-4} \cmidrule(l){5-6} \cmidrule(l){7-8} \cmidrule(l){9-10}
		& N   & TH  & MM & TH & MM  & TH & MM & TH & MM \\ \midrule
		Mild               & 50  & 43.21  & 28.03 & 49.23 & 46.37  & 49.25   & 46.37  & 48.24 & 45.72 \\
		Severe           & 20  & 20.00  & 19.99 & 20.00 & 20.00  & 20.00   & 20.00   & 20.00 & 20.00 \\
		Level              & 20  & 1.07   & 0.14  & 4.26 & 1.17    & 3.98     & 0.83     & 8.19 & 5.61 \\ 
		Catchup         & 10   & 0.27    & 0.00  & 2.05   & 0.60    & 1.67     & 0.38    & 4.72 & 3.83  \\ \midrule
		\textbf{Total} & 100  & 64.55 & 48.17 & 75.54  & 68.14  & 74.90  & 67.58    & 81.15 & 75.16 \\ \bottomrule
	\end{tabular}
	\vskip5mm
	
	\subcaption{Proportion faltering = 20\%.}
	\begin{tabular}{lccccccccc} 
		\toprule
		&  & \multicolumn{2}{c}{SDS} & \multicolumn{2}{c}{RS}  & \multicolumn{2}{c}{ARS} & \multicolumn{2}{c}{MRS} \\    \cmidrule(){3-4} \cmidrule(l){5-6} \cmidrule(l){7-8} \cmidrule(l){9-10}
		& N   & TH  & MM & TH & MM  & TH & MM & TH & MM \\ \midrule
		Mild               & 100  & 92.16  & 68.22   & 99.37  & 96.30   & 99.35   & 95.34  & 97.44 & 95.41 \\
		Severe           & 40   & 40.00  & 40.00   & 40.00  & 40.00   & 40.00   & 39.60   & 40.00 & 40.00 \\
		Level              & 40   & 3.95    & 0.48     & 12.40   & 3.88     & 10.58   & 2.47     & 21.84 & 19.25 \\ 
		Catchup         & 20   & 1.18    & 0.11     & 6.16     & 1.85    & 4.21    & 0.95      & 13.08 & 12.05 \\ \midrule
		\textbf{Total} & 200  & 137.29 & 108.81 & 157.93 & 142.03 & 154.14 & 138.36   & 172.36 & 166.71 \\ \bottomrule
	\end{tabular}
\end{table}

\begin{table}[h]
	\tbl{Agreement between threshold-based (TH) and mixture model-based (MM) classification under various velocity metrics, measured using percentage discordance (\%D), Cohen's kappa ($\kappa$) and its significance test (significance level = $1\%$). The numbers are averages across 100 simulations.}
	{\begin{tabular}{lccccccccc} \toprule
			&   \multicolumn{3}{c}{5\%}  &  \multicolumn{3}{c}{10\%} &  \multicolumn{3}{c}{20\%} \\     \cmidrule(){2-4}  \cmidrule(l){5-7} \cmidrule(l){8-10} 
			Metric  & \%D & $\kappa$ & \%Sig  & \%D & $\kappa$ & \%Sig  & \%D & $\kappa$ & \%Sig \\ \midrule
			SDS & 2.61 & 0.63 & 100 & 4.85 & 0.65 & 100 & 8.32 & 0.69 & 100 \\
			RS   & 1.58  & 0.80 & 100 & 2.88 & 0.82 & 100 & 5.14 & 0.82 & 100 \\
			ARS & 1.59  & 0.80 & 100 & 2.92 & 0.81 & 100 & 5.49 & 0.81 & 100 \\ 
			MRS & 1.26 & 0.85 & 100 & 1.88 & 0.88 & 100 & 1.90 & 0.94 & 100 \\ \bottomrule
	\end{tabular}}
	\label{tab:sim_agreesparse}
\end{table}

\section{Application} \label{sec:application}

\subsection{Data}

The Healthy Birth, Growth and Development knowledge integration (HBGDki) project is a collaboration funded by the Bill and Melinda Gates Foundation, with the aim of improving the overall health and wellbeing of children across the world. While the project includes some studies from countries such as the United States, the majority are from low to middle income countries. So far, the project has gathered data from more than 190 studies representing almost 12 million children from 36 countries.

We used a study from the HBGDki program that was previously analyzed by Leung et al. \cite{Leung2017conditional} using conditional random slopes. The birth cohort from Vellore, India consists of 373 children with anthropometric measurements from birth up to around 3 years old \cite{Paul2014rotavirus}. In this analysis, we restricted our attention to the standardized height, HAZ from birth up to 1 year old, although the methods considered could also be applied to other time periods and anthropometric measures such as weight-for-age z-scores (WAZ), weight-for height z-scores (WHZ), and BMI-for-age z-scores (BAZ). We excluded extreme HAZ values (HAZ $< 6$ or HAZ $> 6$) based on WHO recommendations. 

After taking into account of these exclusion criteria, the resulting dataset consists of $5$ to $15$ observations per child, with the median number of observations being $12$. The age at baseline observation ranges from $1$ to $225$ days, with the median age being $29$ days. Note that day $1$ is considered the day of birth. The age at followup (closest to but not exceeding $1$ year old) ranges from $172$ to $365$ days, with the median being $349$ days. The large age spread at baseline and followup renders SDS unsuitable as it can only be used to compare children measured at the same time points. For this reason, we do not consider SDS-based metrics in the analysis. In practice, it is very common for the timing and frequency of measurements to differ across the children, even though measurements were scheduled regularly. The HAZ at baseline and followup has a median of $-0.8$ $(1.3)$ and $-2.0$ $(1.3)$, respectively, where the numbers in the bracket denote standard deviation. This is common in low-to-middle income countries, where the median HAZ is constantly below $0$ and declines as children grow. Due to the variability of HAZ at baseline, we also consider the conditional velocity metrics which account for regression to the mean (RTM), in addition to the unconditional velocity metrics. 

\subsection{Results}

\begin{figure}[h]
	\centering
	\includegraphics[scale=0.5]{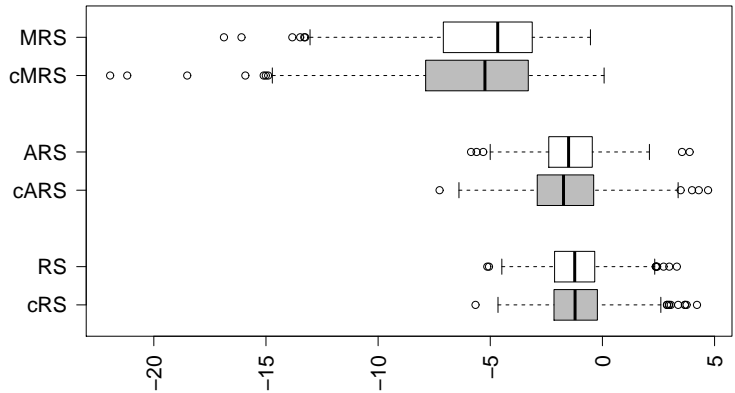} 
	\caption{Velocity estimates of children within the India data set under random slopes (RS), average random slopes (ARS), minimum random slopes (MRS), and their conditional versions (shaded).}  \label{fig:vel_boxplot}
\end{figure}
\begin{figure}[hp]
	\centering
	\begin{subfigure}{\textwidth}
		\centering
		\includegraphics[scale=0.45]{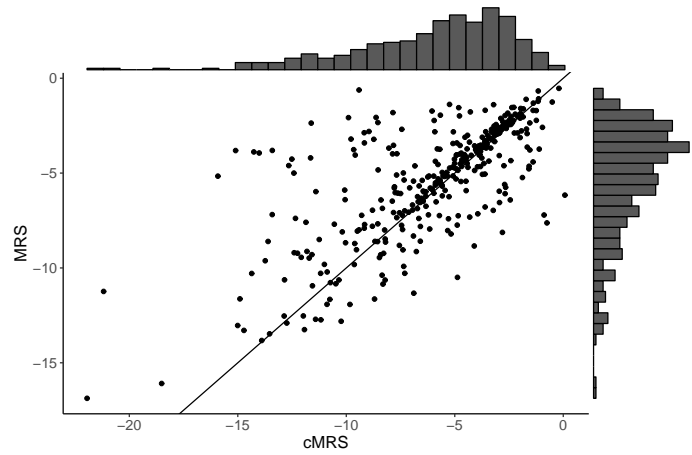}
		\caption{Minimum random slopes versus conditional minimum random slopes.}
		\label{fig:scatter_MRSvsCMRS}
	\end{subfigure}
	\vskip0.35cm
	\begin{subfigure}{\textwidth}
		\centering
		\includegraphics[scale=0.5]{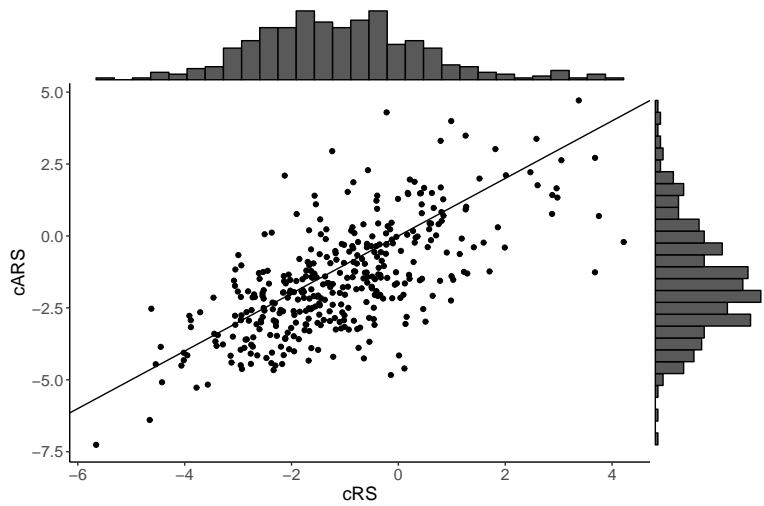}
		\caption{Conditional average random slopes versus conditional random slopes.}
		\label{fig:scatter_CRSvsCARS}
	\end{subfigure}
	\vskip0.35cm
	\begin{subfigure}{\textwidth}
		\centering
		\includegraphics[scale=0.5]{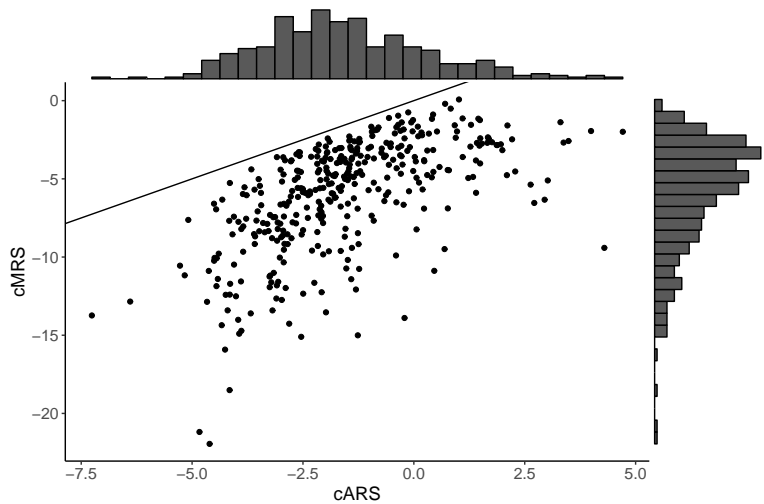}
		\caption{Conditional minimum random slopes versus conditional average random slopes.}
		\label{fig:scatter_CARSvsCMRS}
	\end{subfigure}
	\caption{Comparison of velocity estimates of children within the India data set.}
	\label{fig:vel_scatterplot}
\end{figure}

Figure \ref{fig:vel_boxplot} displays the velocity estimates using the various metrics considered in this article. RS and ARS-based metrics are largely similar, which conforms with the finding from the simulation study (see also Figure \ref{fig:scatter_CRSvsCARS}). As expected, MRS-based metrics produced velocity estimates with much smaller values (see also Figure \ref{fig:scatter_CARSvsCMRS}). The velocity estimates produced by cMRS are generally lower than those of MRS (see also Figure \ref{fig:scatter_MRSvsCMRS}), as cMRS took into account the fact that most children exhibit a HAZ value below $0$ at baseline. These children should have caught up during their first year due to the effect of RTM, but instead their HAZ declined throughout their first year. Under cMRS, children who started with a positive HAZ at baseline need to exhibit a steep decline in order to be classified as having faltered, whereas children who started with a negative HAZ would require a milder decline in HAZ.

\begin{figure}[hp]
	\begin{subfigure}{\textwidth}
		\centering
		\includegraphics[scale=0.4]{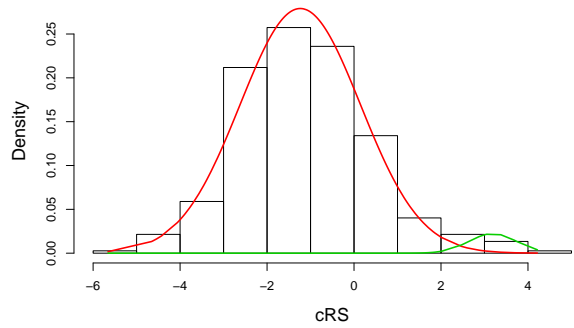}
		\caption{Conditional random slopes.}
		\label{fig:mixture_RS}
	\end{subfigure}
	\begin{subfigure}{\textwidth}
		\centering
		\includegraphics[scale=0.4]{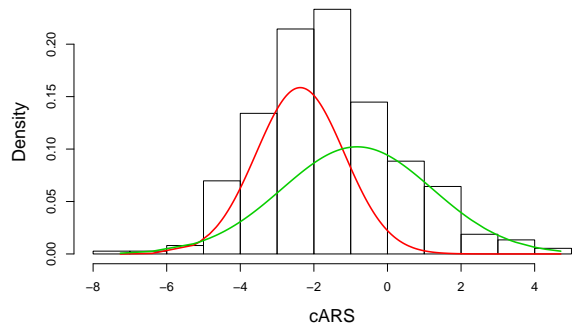}
		\caption{Conditional average random slopes.}
		\label{fig:mixture_ARS}
	\end{subfigure}
	\begin{subfigure}{\textwidth}
	\centering
	\includegraphics[scale=0.4]{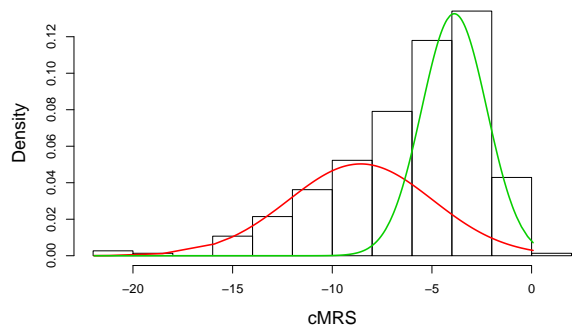}
	\caption{Conditional minimum random slopes.}
	\label{fig:mixture_MRS}
	\end{subfigure}
	\caption{Velocity estimates of children from the India data set, fitted using Gaussian mixture components.}
	\label{fig:vel_mixture}
\end{figure}

\begin{figure}[hp]
	\centering
	\begin{subfigure}{\textwidth}
		\centering
		\includegraphics[scale=0.5]{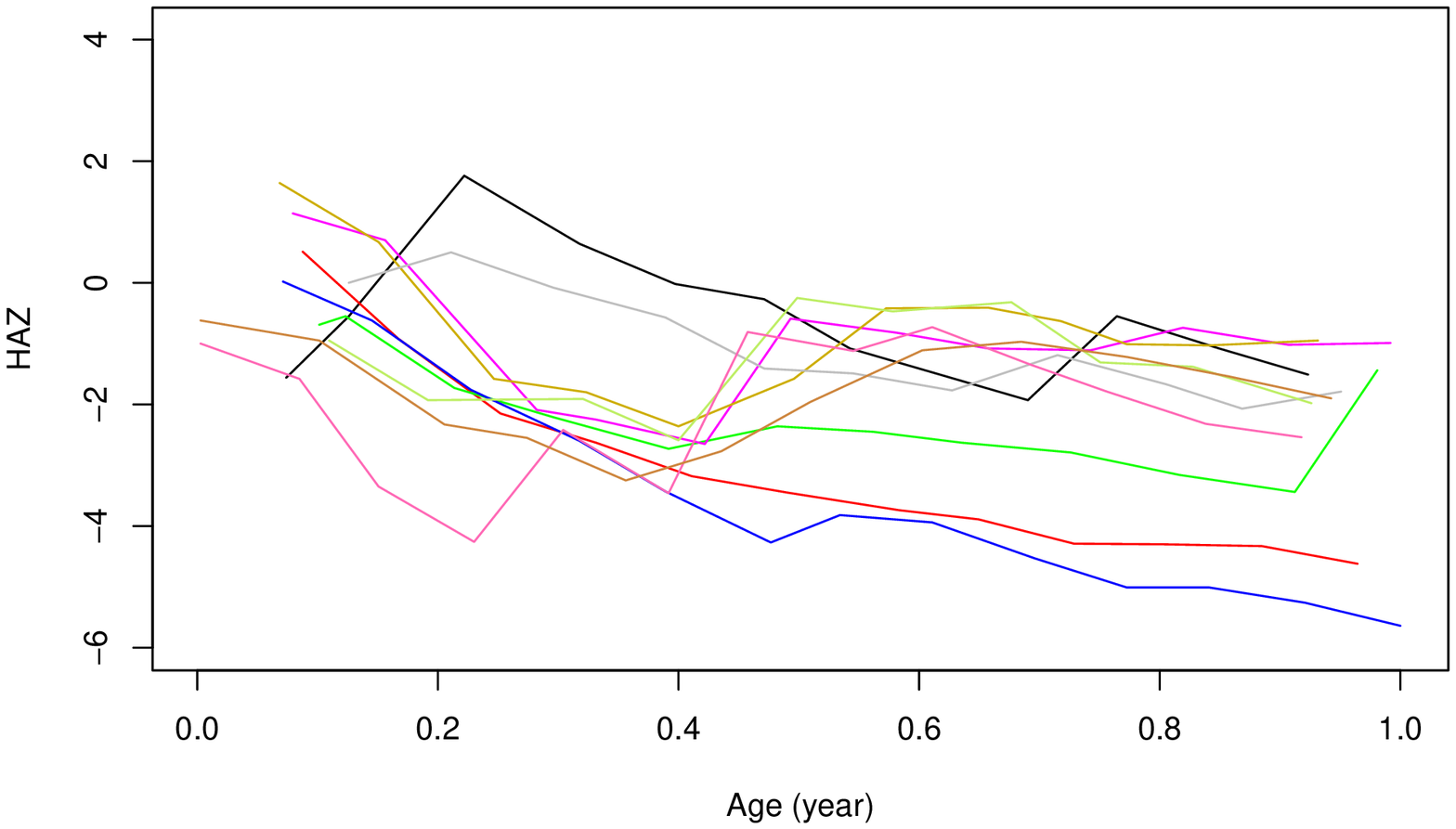}
		\caption{Faltering group.}
		\label{fig:traj_neverfalter}
	\end{subfigure}
	\begin{subfigure}{\textwidth}
		\centering
		\includegraphics[scale=0.5]{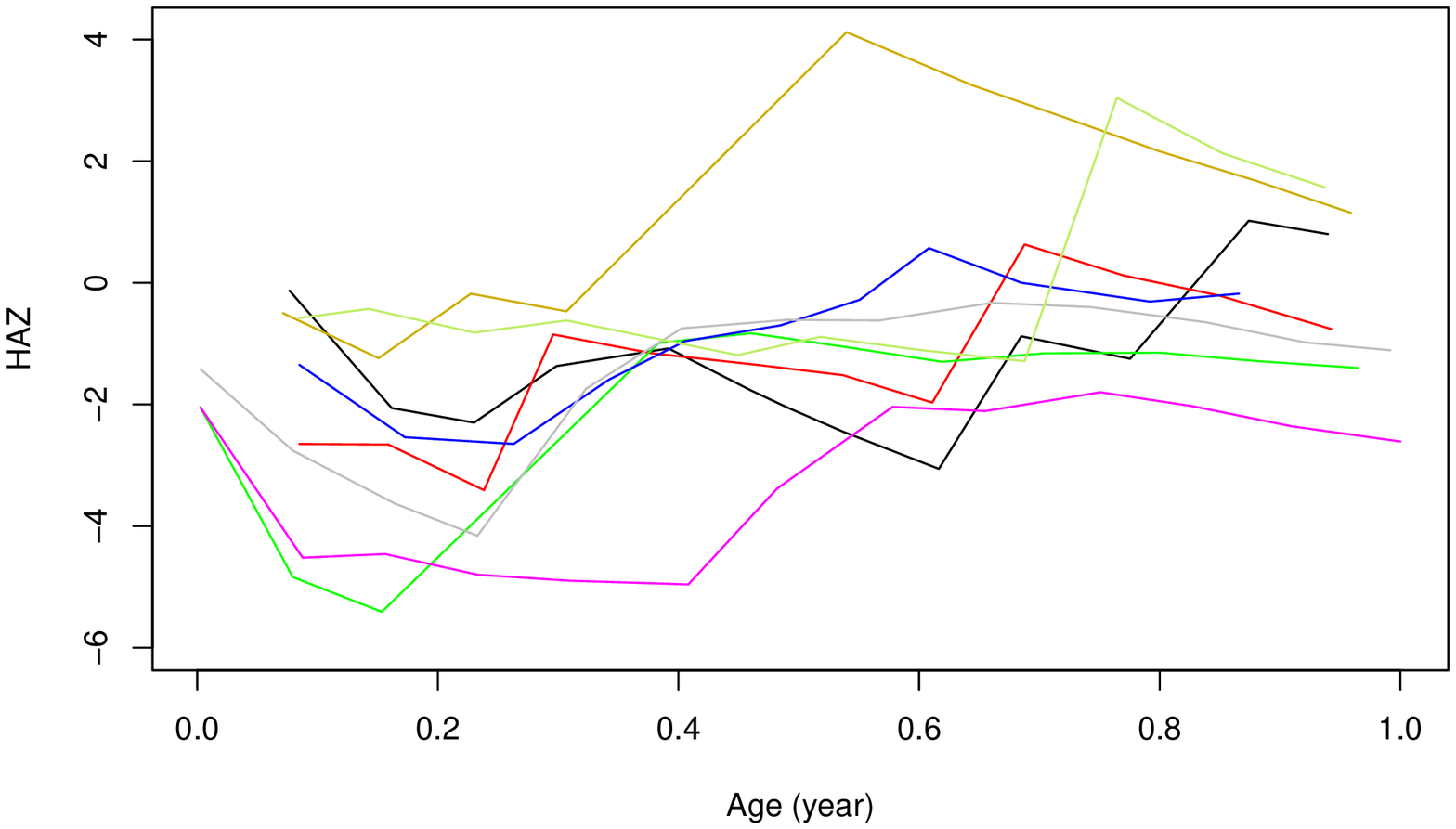}
		\caption{Non-faltering group.}
		\label{fig:traj_falterMRS}
	\end{subfigure}%
	\caption{Raw trajectories of a sample of children from the India data set classified into faltering and non-faltering groups using Gaussian mixture components fitted on conditional minimum random slopes (cMRS).}
	\label{fig:vel_traj}
\end{figure}

Figure \ref{fig:vel_mixture} shows the three conditional velocity estimates fitted using Gaussian mixture components. The histograms for cRS and cARS are not well separated, rendering the mixture model unsuitable. This is not surprising as cRS is based on a linear mixed model, which is not flexible enough to capture the changing velocities throughout the period of growth. Although cARS is based on a piecewise linear mixed model, the metric is calculated by weighting the velocities throughout the period of growth, effectively ``cancelling out" the extreme values. Forcing a mixture model on these metrics resulted in components that are essentially homogeneous (Figures \ref{fig:mixture_RS} \& \ref{fig:mixture_ARS}). On the other hand, mixture models were able to separate the children who have faltered under cMRS (Figure \ref{fig:mixture_MRS}), giving final mixing proportions of 0.46 and 0.54, with corresponding means (standard deviations) of -8.57 (3.62) and -3.88 (1.64). The raw trajectories of a selection of children who were classified into the faltering and non-faltering groups under the cMRS metric are shown in Figure \ref{fig:vel_traj}.  Visual inspection reveals that the MRS-based metric performed well at separating out children who have faltered, when used in conjunction with mixture model-based classification.

\section{Concluding Remarks} \label{sec:remark}
 
In this article, we proposed two new velocity metrics, the average random slopes (ARS) and the minimum random slopes (MRS). We also proposed classifying children into faltering and non-faltering groups using a data-driven approach via mixture model-based (MM) classification. Compared to the ``one-size-fits-all" threshold commonly used in child growth literature that is subject to personal choice, the  MM classification allows the data to guide the choice of the threshold. We showed in a simulation study that MRS, when used in conjunction with MM classification, exhibits superior performance compared to existing methods. The method was subsequently applied to the India data set from the HBGDki database.  

Having said that, faltering children is defined relative to the population under consideration. If we consider a birth cohort from a developed country, it may well be that there are no faltering children from a practical point of view, but those children who had a slower growth compared to the rest would be classified as having faltered. In these scenarios, we would not want to use the proposed velocity metrics to classify children into the faltering versus non-faltering group, but they could still be used as a predictor or response in a regression model. 


The methods presented in this article have several natural extensions. It is possible to built a classifier based on multivariate anthropometric measures (such as HAZ, WAZ, WHZ, and BAZ) using multivariate mixture models \cite{Gage2003classification}.  Alternatively, we could also develop a composite velocity index based on multivariate anthropometric measures and use it for classification. In this article, we extract velocity metrics from fitted growth trajectories, and subsequently use it to built a classifier. An alternative approach would be to incorporate the classification within the modeling framework, using a piecewise linear mixed model with heterogeneity in child-specific random effects, modeled via a mixture of Gaussian distributions. This is more natural compared to our two-step approach, and it also permits the use of multidimensional velocity estimates to build a classifier, as opposed to our approach that summarizes the velocities from multiple segments in the piecewise linear mixed model into a single estimate. Here, we use a cutoff value of 0.5 for classification, but one could also explore the effect of changing the cutoff values on classification accuracy. There are also avenues for methodological work on inference for minimum random slopes. 


\section*{Acknowledgement(s)}
The authors wish to recognize the principal investigators and their study team members for their generous contribution of the data that made this report possible and the members of the Bill \& Melinda Gates Foundation's Knowledge Integration team who directly or indirectly contributed to the study.

\section*{Disclosure statement}

No potential conflict of interest was reported by the authors.

\section*{Funding}

This study was supported by the Bill \& Melinda Gates Foundation via the Healthy Birth, Growth, and Development knowledge integration (HBGDki) project. The article contents are the sole responsibility of the authors and may not necessarily represent the official views of the Bill \& Melinda Gates Foundation or other agencies that may have supported the primary data studies used in the present study.

\section*{Notes on contributors}

\textbf{\textit{Jarod Y. L. Lee}} is a postdoctoral research fellow at the University of Technology Sydney. He received his bachelor's degree in actuarial science from the Australian National University, and his PhD degree in statistics from the University of Technology Sydney.  His research interests included longitudinal and multilevel data analysis, growth modeling, classification methods, and more generally applied statistics. He received a teaching award from the University of New South Wales in 2015, and has consulted by the Australian government on a copyright survey project.  \\

\noindent \textbf{\textit{Craig Anderson}} is a lecturer in statistics at the University of Glasgow. He worked at the University of Technology Sydney as a postdoctoral research fellow, where he contributed to a variety of research projects related to child growth modeling as part of the Bill \& Melinda Gates Foundation’s HBGDki project. \\

\noindent \textbf{\textit{Wai T. Hung}} is a postdoctoral research fellow at the University of Technology Sydney. He attained his PhD degree in statistics from Macquarie University in 1989 and  has extensive work experience in public health and market research. His research interests are meta-analysis of big health data and computer-aided diagnosis. \\

\noindent \textbf{\textit{Hon Hwang}} is a PhD candidate at the University of Technology Sydney under the supervision of Professor Louise Ryan. He attained his undergraduate degree in engineering from the University of Technology Sydney, and worked as a research engineer at Australia's national scientific research organization, CSIRO. \\

\noindent \textbf{\textit{Louise M. Ryan}} is a distinguished professor of statistics at the University of Technology Sydney and an adjunct professor of biostatistics at the Harvard University. She is a chief investigator of the Australian Research Council Centre of Excellence for Mathematical and Statistical Frontiers, and holds a grant from the Bill and Melinda Gates Foundation. Louise is the president of the International Biometric Society and served in a variety of professional capacities, including editor-in-chief of Statistics in Medicine, co-editor of Biometrics and President of the Eastern North American Region of the International Biometric Society. She has received numerous prestigious awards, including her 2012 election to the Australian Academy of Science, a 2015 honorary doctorate from Ghent University, Harvard's 2015 Centennial Medal, and Pitman Medal in 2018. She has published over 300 articles in peer-reviewed journals, and is well known for her methodological contributions to statistical methods for cancer and environmental health research. \\
  

\bibliographystyle{tfnlm} 
\bibliography{hbgdki} 

%
%

\end{document}